\documentclass{cjpsuppl}
\usepackage{graphics} 
\usepackage{epsfig}
\usepackage{wrapfig}
\newcommand{\la}{\langle}
\newcommand{\ra}{\rangle}
\newcommand{\di}{ {\rm d} }

\newcommand{\ph}{\phantom{X}}
\newcommand{\SIDIS}{{\mbox{\tiny SIDIS}}}
\newcommand{\DY   }{{\mbox{\tiny DY}}}
\newcommand{\aH}{ {a} }
\begin{document}
 \slacs{.6mm}
 \title{Status of Sivers and Collins Single Spin Asymmetries}
 \authori{}
 \addressi{}
 \authorii{\underline{A.~V.~Efremov}}
 \addressii{Joint Institute for Nuclear Research, Dubna, 141980 Russia}
 \authoriii{K.~Goeke and P.~Schweitzer}
 \addressiii{Institut f\"ur Theoretische Physik II, Ruhr-Universit\"at
Bochum, Germany}
 \authoriv{}
 \addressiv{}
 \authorv{}     \addressv{}
 \authorvi{}    \addressvi{}
 \headtitle{Status of Sivers and Collins SSA }
 \headauthor{A.~V.~Efremov \it at al.}
 \lastevenhead{A.~V.~Efremov {\it at al.}, Status of Sivers and 
 Collins SSA}
 \refnum{}
 \daterec{..., 2006} 
 \suppl{A}  \year{2007}
 \setcounter{page}{1}
\maketitle

\begin{abstract}
The Sivers and Collins asymmetries are 
the most interesting Single Spin Asymmetries in semi-inclusive 
deeply inelastic scattering with transverse target polarization. 
In this talk we present our understanding of these phenomena.
\end{abstract}

\section{Introduction}

Single spin asymmetries (SSA) in hard reactions have a long 
history dating back to the 1970s when significant polarizations of 
$\Lambda$-hyperons in collisions of unpolarized hadrons were 
observed \cite{Bunce:1976yb}, and to the early 1990s when large 
asymmetries in $p^\uparrow p\to\pi X$ or $p^\uparrow \bar{p}\to\pi 
X$ were found at FNAL \cite{Adams:1991rw}. No fully consistent and 
satisfactory unifying approach to the theoretical description of 
these observations has been found so far --- see the reviews 
\cite{Felix:1999tf,Anselmino:2002mx}.

Interestingly, the most recently observed SSA phenomena, namely
those in semi-inclusive deeply inelastic scattering (SIDIS) seem
better under control. This is in particular the case for the
transverse target SSA observed at HERMES and COMPASS
\cite{Airapetian:2004tw,Alexakhin:2005iw,Diefenthaler:2005gx}. On
the basis of a generalized factorization approach in which
transverse parton momenta are taken into account
\cite{Collins:1981uk,Ji:2004wu,Collins:2004nx} these ``leading
twist'' asymmetries can be explained \cite{Boer:1997nt} in terms
of the Sivers
\cite{Sivers:1989cc,Brodsky:2002cx,Collins:2002kn,Belitsky:2002sm}
or Collins effect \cite{Collins:1992kk}. The former describes,
loosely speaking, the distribution of unpolarized partons in a
transversely polarized proton, the latter describes the
fragmentation of transversely polarized partons into unpolarized
hadrons. In the transverse target SSA these effects can be
distinguished by the different azimuthal angle distribution of the
produced hadrons: Sivers  effect $\propto\sin(\phi-\phi_S)$, while
Collins effect $\propto\sin(\phi+\phi_S)$, where $\phi$ and
$\phi_S$ denote respectively the azimuthal angles of the produced
hadron and the target polarization vector with respect to the axis
defined by the hard virtual photon \cite{Boer:1997nt}. Both
effects have been subject to intensive phenomenological studies in
hadron-hadron-collisions
\cite{Anselmino:1994tv,Anselmino:1998yz,D'Alesio:2004up,Anselmino:2004ky,Ma:2004tr}
and in SIDIS
\cite{Efremov:2004tp,Anselmino:2005nn,Anselmino:2005ea,Vogelsang:2005cs}.
For the longitudinal target SSA in SIDIS, which were observed
first but are dominated by subleading-twist effects, the situation
is less clear and their description ({\sl presuming} 
that factorization holds) is more involved.

In this talk our understanding of these phenomena is presented.

\section{Sivers effect}
The Sivers effect \cite{Sivers:1989cc} was originally suggested
to explain the large single spin asymmetries (SSA) observed in
$p^\uparrow p\to\pi X$ (and $\bar{p}^\uparrow p\to\pi X$) at FNAL
\cite{Adams:1991rw} and recently at higher energies in the RHIC
experiment \cite{Adams:2003fx}. The effect considers a
non-trivial correlation between (the transverse component of) the
nucleon spin ${\bf S}_{\rm T}$ and intrinsic transverse parton
momenta ${\bf p}_{\rm T}$ in the nucleon. It is proportional to
the ``(naively or artificially) T-odd'' structure $({\bf S}_{\rm
T}\times{\bf p}_{\rm T}){\bf P}_N$ and quantified in terms of the
Sivers function $f_{1T}^\perp(x,{\bf p}_T^2)$
\cite{Kotzinian:1995,Boer:1997nt} responsible for the left-right
asymmetry in transversaly polarized nucleon whose precise
definition in QCD was worked out only recently
\cite{Brodsky:2002cx,Collins:2002kn,Belitsky:2002sm}.

Our approach to the description of the Sivers effect in SIDIS and 
Drell-Yan was described in detail on the spin conference in 
Prague last year \cite{Efremov-Praha05} (see also 
\cite{Efremov:2004tp,Collins:2005ie,Collins:2005rq}), and we will 
restrict ourselves to review here only the main points.

\subsection{Sivers effect in SIDIS}

\begin{wrapfigure}{R}{.41\textwidth}
\begin{center}
\vspace{-3mm}
\includegraphics[width=.38\textwidth]
{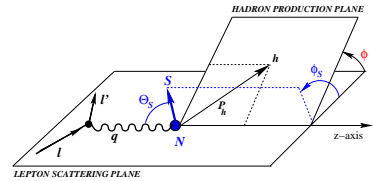}
\end{center}

\vspace{-7mm}\hfil
\begin{minipage}{.35\textwidth}
\caption{\label{fig2-processes-kinematics}\footnotesize 
Kinematics of the SIDIS process $lp\rightarrow l^\prime h X$. }
\end{minipage}
\end{wrapfigure}

The azimuthal SSA measured by HERMES \& COMPASS in the SIDIS 
process $lp^\uparrow\rightarrow l'h X$ (see 
Fig.~\ref{fig2-processes-kinematics}) is defined as
\be
\frac{N^\uparrow -N^\downarrow}{N^\uparrow+ N^\downarrow}\propto
\underbrace{\dots\;\sin(\phi-\phi_S)}_{\hspace{1cm} \mbox{Sivers
\hspace{0.3cm} and}\hspace{-1.5cm}} +
\underbrace{\dots\;\sin(\phi+\phi_S)}_{\hspace{1cm}
\mbox{Collins effect}}
\ee

\noindent 
where $N^{\uparrow(\downarrow)}$ are the event counts 
for the respective transverse target  polarization. We assume the 
distributions of transverse parton and hadron momenta in 
distribution (DF) and fragmentation function (FF) to be Gaussian 
with corresponding averaged transverse momenta to be flavour and 
$x$- or $z$-independent. The experimental data 
\cite{Airapetian:2004tw,Alexakhin:2005iw} are presented in the 
form of \cite{Collins:2005ie}
\bea
\label{Eq:AUT-SIDIS-Gauss}
        A_{UT}^{\sin(\phi-\phi_S)} &=& (-2)\; \frac{a_{\rm Gauss}
        \sum_a e_a^2\,x f_{1T}^{\perp(1)a}(x)\, D_1^{a}(z)}{
        \sum_a e_a^2\,x f_1^a(x)\,D_1^{a}(z)}\\\nonumber
        \mbox{with}\;\;\;
        a_{\rm Gauss} &=& \frac{\sqrt{\pi}}{2}\;
        \frac{M_N}{\sqrt{p^2_{\rm Siv}+K^2_{\! D_1}/z^2}}
\eea
and with
\be
\label{Eq:Def-Siv-transverse-mom}
        f_{1T}^{\perp(1)a}(x)
        \equiv \int\!\di^2{\bf p}_T\; \frac{{\bf p}_T^2}{2 M_N^2}\;
                   f_{1T}^{\perp a}(x,{\bf p}_T^2)
        \stackrel{\rm Gauss}{=}
        \frac{\la p_T^2\ra_{\rm Siv}^{\phantom{2}}}{2 M_N^2}\;
        f_{1T}^{\perp a}(x) \;.
\ee

We use predictions from the QCD limit of a large number of colours
$N_c$. In this limit it was proven in a model independent way that
\cite{Pobylitsa:2003ty}
\be\label{Eq:large-Nc}
      f_{1T}^{\perp u}(x,{\bf p}_T^2) =
    - f_{1T}^{\perp d}(x,{\bf p}_T^2) \;\;\;
    \mbox{modulo $1/N_c$ corrections,}\ee
for not too small and not too large $x$ satisfying
$xN_c={\cal O}(N_c^0)$.
Analog relations hold for the Sivers antiquark distributions.

\begin{figure}
\begin{tabular}{ccc}
\includegraphics[width=0.3\textwidth]{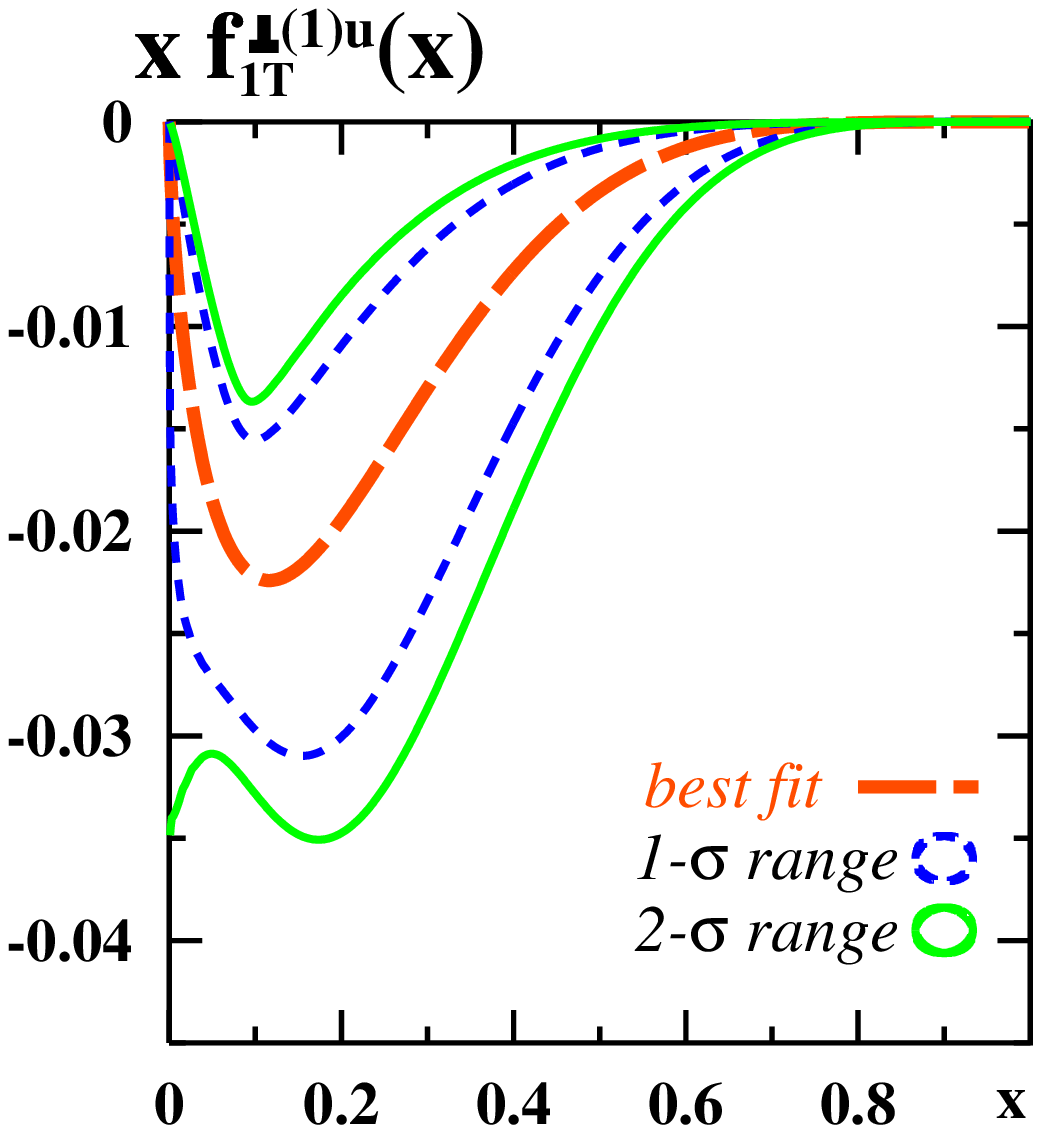}&
\includegraphics[width=0.3\textwidth]{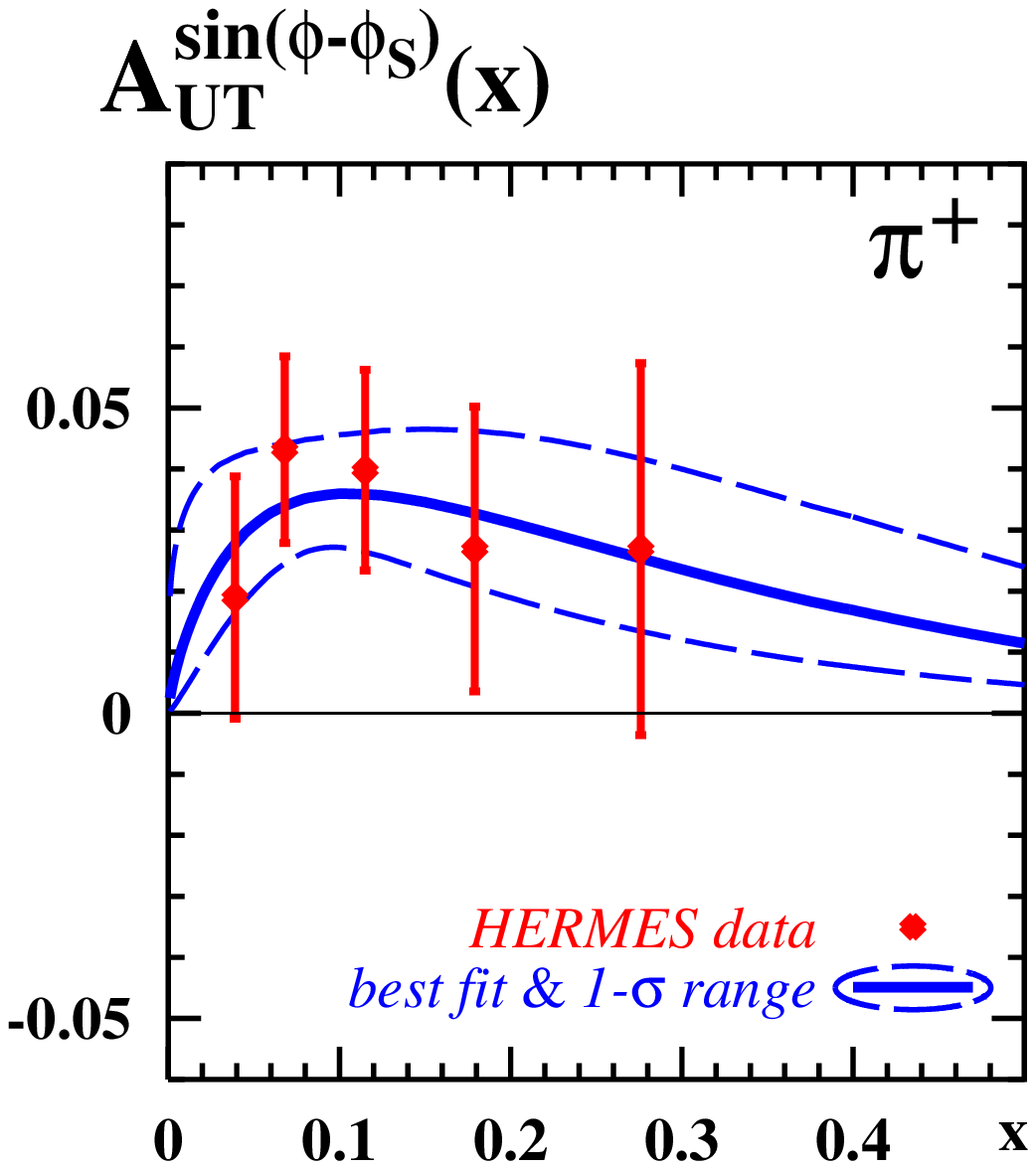}&
\includegraphics[width=0.3\textwidth]{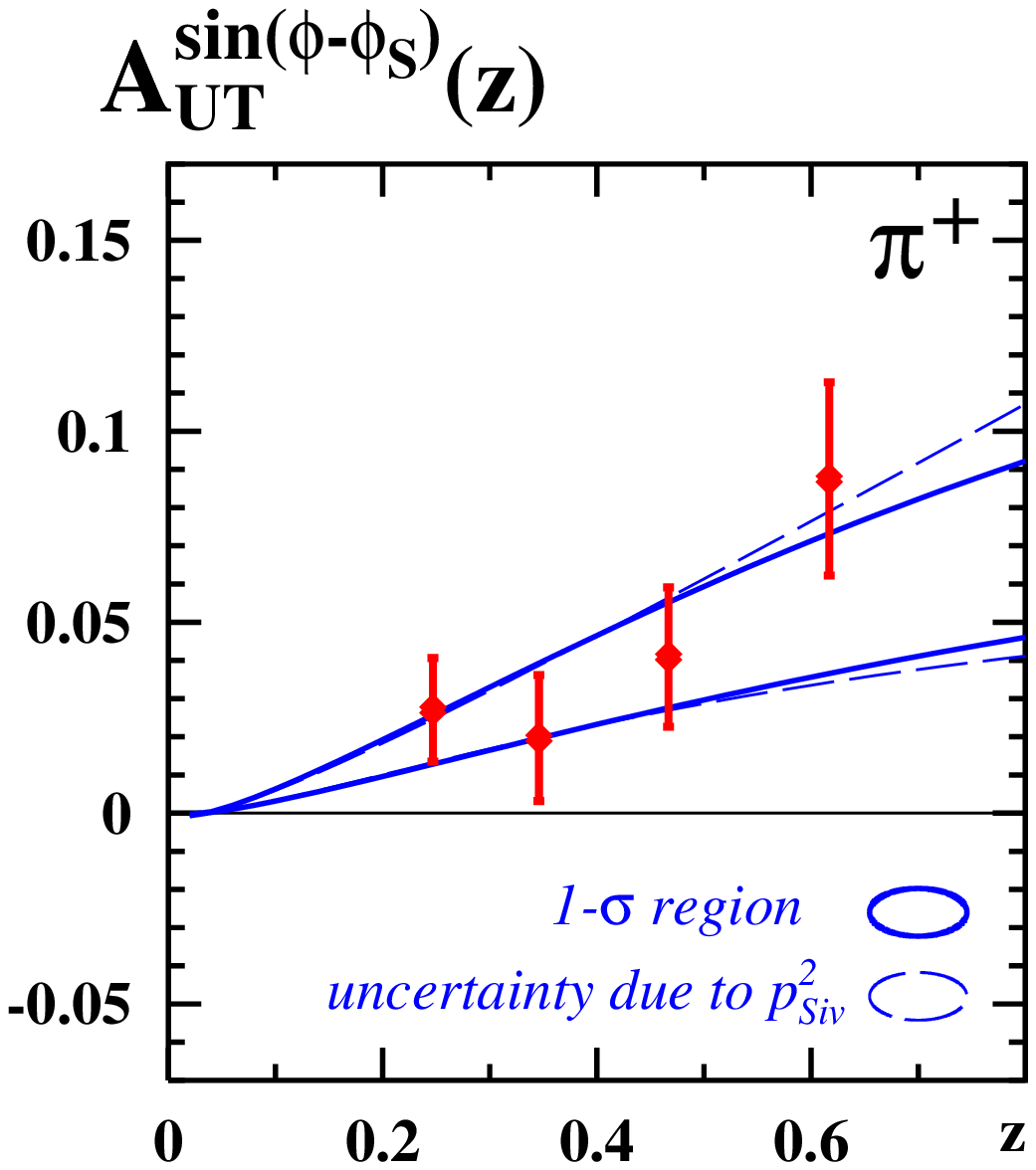}\\
{\bf a}&{\bf b}&{\bf c}
\end{tabular}

\vspace{-5mm}
\caption{\label{Fig5-compare-to-data}\footnotesize
        {\bf a.} The $u$-quark Sivers function
        $xf_{1T\SIDIS}^{\perp(1)u}(x)$ vs.\  $x$ at a scale of
        about $2.5\,{\rm GeV}^2$, as obtained from a fit to the
        HERMES data \cite{Airapetian:2004tw}. Shown are the best
        fit and its 1- and 2-$\sigma$ uncertainty.
        \newline
        {\bf b.} and {\bf c.} The azimuthal SSA
        $A_{UT}^{\sin(\phi_h-\phi_S)}$ as function of $x$ and $z$
        for positive pions as obtained from the fit
        (\ref{Eq:ansatz+fit}) in comparison to the {\sl final}
        HERMES data \cite{Airapetian:2004tw}.}
\end{figure}

Imposing the large-$N_c$ relation (\ref{Eq:large-Nc}) as an
additional constraint, and neglecting effects of antiquarks and
heavier flavours, it is shown \cite{Collins:2005ie} that the
HERMES data \cite{Airapetian:2004tw} can be described by the
following 2-parameter ansatz and best fit
\be\label{Eq:ansatz+fit}
        f_{1T\SIDIS}^{\perp (1) u}(x) = -f_{1T\SIDIS}^{\perp (1) d}(x)
        \stackrel{\rm ansatz}{=}\;  A \, x^b   \,(1-x)^5
        \,\,\stackrel{\rm fit}{=}\,\, -0.17 x^{0.66}(1-x)^5\;.
\ee
The fit and its 1- and 2-$\sigma$ uncertainty due to the
statistical error of the data \cite{Airapetian:2004tw} are shown
in Fig.~\ref{Fig5-compare-to-data}a. In
Fig.~\ref{Fig5-compare-to-data}b and c we compare the Sivers SSA
obtained on the basis of our fit (\ref{Eq:ansatz+fit}) to the
HERMES data \cite{Airapetian:2004tw}.

%
%

As an intermediate summary we conclude that the HERMES data
\cite{Airapetian:2004tw} are well compatible with the large-$N_c$
predictions (\ref{Eq:large-Nc}) for the Sivers function
\cite{Pobylitsa:2003ty} and that the fit (\ref{Eq:ansatz+fit})
satisfies the positivity bounds \cite{Bacchetta:1999kz}.
Remarkably, the sign of the extracted Sivers function in
Eq.~(\ref{Eq:ansatz+fit}) agrees with the physical picture
discussed in \cite{Burkardt:2002ks}.

In our fitting procedure we did not use the HERMES data
\cite{Airapetian:2004tw} on the $z$-dependence of the Sivers SSA.
These data could have been used as an additional constraint for
the integrals of $x\,f_{1T}^{\perp (1) a}(x)$ in the range $0.023
< x < 0.4$, which corresponds to the cuts in the HERMES
experiment. This would have helped to improve the significance of
the fit, considering that only few $x$-data points are available.
Instead, we use these data as a valuable cross check of our
approach. As the $z$-shape of the SSA is dictated by the
unpolarized fragmentation function $D_1^a(z)$ {\sl and} the
$z$-dependence of the Gaussian factor $a_{\rm Gauss}$ in
Eq.~(\ref{Eq:AUT-SIDIS-Gauss}), this is not only a cross check
for the extracted Sivers function, but it also
tests the Gauss and the large-$N_c$
ansatz (\ref{Eq:large-Nc}) itself. In
Fig.~\ref{Fig5-compare-to-data}c we confront our fit result
(\ref{Eq:ansatz+fit}) with the $z$-dependent HERMES data on the
Sivers SSA \cite{Airapetian:2004tw}. We observe a satisfactory
agreement.


In the ansatz (\ref{Eq:ansatz+fit}) we neglected the Sivers
distributions for antiquarks and for the strange and heavier
quarks. Are these reasonable approximations?

We have explicitly checked that Sivers $\bar u$- and $\bar 
d$-distributions play a negligible role for the $\pi^+$ SSA, and 
give more pronounced effects for the $\pi^-$ SSA. In fact, we 
found that even sizeable Sivers  $\bar u$- and $\bar 
d$-distributions cannot be resolved within the error bars of the 
present data \cite{Airapetian:2004tw}. Also the neglect of Siver 
strangeness seems reasonable. We shall, however, come back to 
this point in the next Section.

Next let us address the $1/N_c$-corrections. In order to have an
idea of the effect of these corrections, let us assume that the
flavour singlet Sivers distribution is not exactly zero but
suppressed by exactly a factor of $1/N_c$ with respect to the
flavour non-singlet combination according to
Eq.~(\ref{Eq:large-Nc}). That is,
\be\label{Eq:model-Nc-corr}
\Big |(f_{1T}^{\perp(1)u}+f_{1T}^{\perp(1)d})(x) \Big | 
\stackrel{!}{=} \pm \;\frac{1}{N_c} 
(f_{1T}^{\perp(1)u}-f_{1T}^{\perp(1)d})(x) \,,
\ee
%
\begin{wrapfigure}[16]{HR}{.35\textwidth}
\vspace{-3mm}
\includegraphics[width=.35\textwidth]{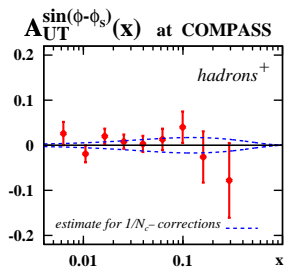}

\vspace{-5mm}\hfil
\begin{minipage}{.3\textwidth}
\caption{\label{Fig9-AUT-Nc-COMPASS}\footnotesize
        The Sivers SSA for $\pi^+$ at COMPASS. The theoretical curves indicate the
        order of magnitude of the effect according to the rough model
        for $1/N_c$-corrections in Eq.~(\ref{Eq:model-Nc-corr}).}
\end{minipage}
\end{wrapfigure}
\noindent 
where we use $f_{1T}^{\perp(1)q}(x)$ from 
Eq.~(\ref{Eq:ansatz+fit}) and set $N_c=3$. 

On a deuteron target, the leading $1/N_c$ prediction gives zero
for the SSA, so that the $1/N_c$ corrections are all that remain.
Assuming for simplicity that the positive and negative hadrons
identified at COMPASS are mainly pions, we obtain in our rough
model (\ref{Eq:model-Nc-corr}) for $1/N_c$-corrections the
results shown in Fig.~\ref{Fig9-AUT-Nc-COMPASS}. Clearly, we see
that the COMPASS data \cite{Alexakhin:2005iw} are compatible with
the large-$N_c$ corrections being of a magnitude compatible with
out model.

Thus, the reason why our large-$N_c$ approach works here, is due
to the fact that current precision of the first experimental data
\cite{Airapetian:2004tw,Alexakhin:2005iw} is comparable to the
theoretical accuracy of the large-$N_c$ relation
(\ref{Eq:large-Nc}). In future, with increasing precision of the
data, it will certainly be necessary to refine the fit ansatz
(\ref{Eq:ansatz+fit}) to include $1/N_c$ corrections and
antiquarks.

Our fit is in qualitative agreement with extractions of the
Sivers function
\cite{Anselmino:2005nn,Anselmino:2005ea,Vogelsang:2005cs} from
the same \cite{Airapetian:2004tw} and from the more recent and
more precise (but {\sl preliminary}) HERMES data
\cite{Diefenthaler:2005gx} (see \cite{Anselmino:2005an} for a
detailed comparison).

\subsection{Sivers effect for kaons}

In the HERMES experiment the RICH detector allows to separate 
kaons and pions. The outcome of the HERMES experiment is as 
follows \cite{Diefenthaler:2006vn}. The Sivers SSA for $K^+$ is 
about (10-15)\% in the region of $x = (0.05-0.15)$, i.e. about 
factor (2-3) larger than the $\pi^+$ SSA. In the region of $x\geq 
0.15$ it is (within large error bars) of comparable size as the 
$\pi^+$ SSA. The $K^-$ Sivers SSA is compatible with zero within 
large error bars. Can one understand this behaviour?

\begin{wrapfigure}[14]{R}{.35\textwidth}
\includegraphics[width=.33\textwidth]{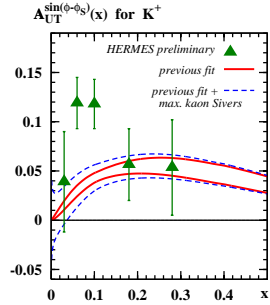}

\vspace{-6mm}\hfil
\begin{minipage}{.3\textwidth}
\caption{\label{fig4kaon}\footnotesize
Estimates for kaon Sivers SSA from a proton target in the
kinematics of the HERMES experiment made on the basis the
understanding of the pion Sivers effect.}
\end{minipage}
\end{wrapfigure}

Let us use the results discussed in Sec 2.1 in order to
estimate the Sivers SSA for kaons at HERMES. For $f^a_1(x)$ we use
the parameterization \cite{Gluck:1998xa}, for the kaon FF the
parameterization \cite{Kretzer:2001pz}. We obtain the results
shown as solid lines in Fig. \ref{fig4kaon}.

In our estimate we have neglected the Sivers effect for strange
quarks. Could therefore the unknown strangeness Sivers function
yield to some surprises and change drastically the picture in Fig.
\ref{fig4kaon} (solid lines)?
%

Let us assume that the $s$- and $\bar s$-Sivers distributions 
saturate the positivity bound \cite{Bacchetta:1999kz} with the 
unpolarized distributions $f^s_1(x)=f^{\bar s}_1$ from 
\cite{Gluck:1998xa}. Then we obtain the result shown as dashed 
line in Fig. \ref{fig4kaon}. We observe that corrections due to 
the strangeness Sivers functions have little impact.

``The only difference'' between the SSA for $\pi^+$ and $K^+$ is 
the exchange $\bar d\leftrightarrow \bar s$. For $R\equiv 
A(K^+)/A(\pi^+)$, i.e.\  the ratio of the $K^+$ to $\pi^+$ Sivers 
SSA, one obtains numerically in the HERMES kinematics 
\be
R=\frac{A(K^+)}{A(\pi^+)}\approx \frac{B(x)+0.35f^{\perp\bar
s}_{1T}(x)}{B(x)+0.09f^{\perp\bar d}_{1T}(x)}\;,
\label{ktopiratio}
\ee
$B(x)\approx f^{\perp u}_{1T}+0.15(f^{\perp
d}_{1T}+4f^{\perp\bar u}_{1T}+ f^{\perp\bar d}_{1T}+f^{\perp
s}_{1T}+f^{\perp\bar s}_{1T})$,
where we used the parameterizations \cite{Gluck:1998xa,Kretzer:2001pz}.

How much Sivers-$\bar q$ is needed to explain $K^+/\pi^+$?
Just to have a better feeling on that, let us consider two models
motivated by our \cite{Efremov:2004tp} and the works
\cite{Anselmino:2005nn,Anselmino:2005ea,Vogelsang:2005cs}:
\begin{itemize}
\item
Model I: \ph $f_{1T}^{\perp Q}\equiv f_{1T}^{\perp u}\approx
-f_{1T}^{\perp d}$,\ph\ph $f_{1T}^{\perp A}\equiv
f_{1T}^{\perp\bar u}\approx f_{1T}^{\perp\bar d} \approx
f_{1T}^{\perp s}\approx -f_{1T}^{\perp\bar s}$
\item
Model II: $f_{1T}^{\perp Q}\equiv f_{1T}^{\perp
u}\approx-2f_{1T}^{\perp d}$,\ph\ph $f_{1T}^{\perp A}\equiv$ same
as above.
\end{itemize}

\begin{wrapfigure}[12]{R}{.35\textwidth}
\vspace{-7mm}
\includegraphics[width=.33\textwidth]{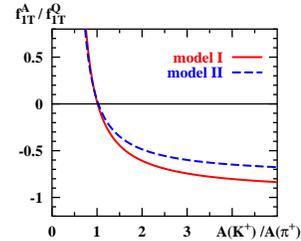}

\vspace{-6mm}\hfil
\begin{minipage}{.3\textwidth}
\caption{\label{fig-R(K/pi)}\footnotesize
The ratio $f_{1T}^{\perp Q}/f_{1T}^{\perp A}$ needed in the
models I \& II in order to explain a given (observed) ratio
$A(K^+)/A(\pi^+)$.}
\end{minipage}
\end{wrapfigure}
With models like I and II the ratio of $K^+$ to $\pi^+$ Sivers SSA
is a function of the variable $\beta=f_{1T}^{\perp
Q}/f_{1T}^{\perp A}$ only. Thus, it is possible to invert this
relation to have $\beta$ as a function of the (measured) $K^+$ to
$\pi+$ Sivers SSA ratio. The exact result of this operation is
shown in Fig. \ref{fig-R(K/pi)} which illustrates that one can
obtain rather large $K^+/\pi^+$ enhancements with still moderate
values of $f_{1T}^{\perp Q}/f_{1T}^{\perp A}$ in the models I and II. 

Of course, the models I and II are particular {\sl choices}. 
A simultaneous refitting will give us a conclusive answer.

\subsection{Sivers effect in the Drell-Yan process}

A particularly interesting feature of the Sivers function (and other
``T-odd'' distributions) concerns the universality property. On the
basis of time-reversal arguments it was predicted
\cite{Collins:2002kn} that $f_{1T}^\perp$ in SIDIS and DY have
opposite sign (we follow the Trento Conventions \cite{Bacchetta:2004jz}),
\be\label{Eq:01}
     f_{1T}^\perp(x,{\bf p}_T^2)_\SIDIS =
    -f_{1T}^\perp(x,{\bf p}_T^2)_\DY \;. \ee The experimental check
of Eq.~(\ref{Eq:01})  would provide a thorough test of our understanding of
the Sivers effect within QCD. In particular, the experimental
verification of (\ref{Eq:01}) is a crucial prerequisite for testing
the factorization approach to the description of processes containing
$p_T$-dependent correlators
\cite{Collins:1981uk,Ji:2004wu,Collins:2004nx}.

On the basis of the fist information of the Sivers effect in SIDIS
\cite{Airapetian:2004tw,Alexakhin:2005iw} it was shown that the 
Sivers effect leads to sizeable SSA in $p^\uparrow\pi^-\to 
l^+l^-X$, which could be studied at COMPASS, and in 
$p^\uparrow\bar{p}\to l^+l^-X$ or $p\bar{p}^\uparrow\to l^+l^-X$ 
in the planned PAX experiment at GSI \cite{PAX,PAX-estimates}, 
making the experimental check of Eq.~(\ref{Eq:01}) feasible and 
promising \cite{Efremov:2004tp}. Both experiments, which could be 
performed in the medium or long term, have the advantage of being 
dominated by annihilation of valence quarks (from $p$) and 
valence antiquarks (from $\bar{p}$ or $\pi^-$), which yields 
sizeable counting rates. Moreover, the processes are not very 
sensitive to the Sivers antiquark distributions, which are not 
constrained by the present SIDIS data, see 
\cite{Efremov:2004tp,Anselmino:2005nn, 
Anselmino:2005ea,Vogelsang:2005cs,Collins:2005ie}.

On a shorter term the Sivers effect in DY can be studied in
$p^\uparrow p\to l^+l^-X$ at RHIC. In $pp$-collisions inevitably 
antiquark distributions are involved, and the counting rates are 
smaller. We have shown, however, that the Sivers SSA in DY can 
nevertheless be measured at RHIC with an accuracy sufficient to 
unambiguously test Eq.~(\ref{Eq:01}) \cite{Collins:2005rq}. We 
refer to 
\cite{Bunce:2000uv,Boer:1999mm,Boer:2003tx,Dressler:1999zv, 
Boer:2001tx,Anselmino:2002pd,Anselmino:2004nk} for discussions of 
further RHIC spin physics prospects, and 
\cite{Boros:1993ps,Hammon:1996pw} for early predictions of SSAs 
in DY.

It remains to be noted that the theoretical understanding of SSA in
$p^\uparrow p\to\pi X$, which originally motivated the introduction
of the Sivers effect, is more involved and less lucid compared to
SIDIS or DY, as here also other mechanisms such as the Collins effect
\cite{Collins:1992kk} and/or dynamical twist-3 effects
\cite{Efremov:eb,Kanazawa:2000hz} could generate SSA.
Phenomenological studies indicate, however, that in a picture based
on $p_T$-dependent correlators the data
\cite{Adams:1991rw,Adams:2003fx} can be explained in terms of the
Sivers effect alone
\cite{Anselmino:1994tv,Anselmino:1998yz,D'Alesio:2004up} with the
other effects playing a less important role
\cite{Anselmino:2004ky,Ma:2004tr}. For recent discussions of
hadron-hadron collisions with more complicated final states (like,
e.g., $p^\uparrow p\to\mbox{jet}_1\mbox{jet}_2 X$) we refer to Refs.\
\cite{Bomhof:2004aw}.


\section{Collins effect}

The chirally odd transversity distribution function $h_1^a(x)$
cannot be extracted from data on semi-inclusive deep inelastic
scattering (SIDIS) alone. It enters the expression for the Collins
single spin asymmetry (SSA) in SIDIS together with the chirally
odd and equally unknown Collins fragmentation function
\cite{Collins:1992kk} (FF) $H_1^a(z)$\footnote{
    \label{Footnote-1}
    We assume a factorized Gaussian dependence on parton and
    hadron transverse momenta \cite{Mulders:1995dh} with
    $B_{\rm G}(z)=(1+z^2\;\la{\bf p}_{h_1}^2\ra/\la{\bf K}^2_{H_1}\ra)^{-1/2}$
    and define $H_1^\aH(z) \equiv H_1^{\perp (1/2) a}(z)= $
    $\int\di^2{\bf K}_T\frac{|{\bf K}_T|}{2zm_\pi} H_1^{\perp a}(z,{\bf K}_T)$
    for brevity.
    The Gaussian widths are assumed flavor and $x$- or $z$-independent.
    We neglect throughout soft factors \cite{Ji:2004wu}.}
\be\label{Eq:AUT-Collins-1}
    A_{UT}^{\sin(\phi+\phi_S)}
    \!= 2\frac{\sum_a e_a^2 x h_1^a(x)B_{\rm G}
    H_1^{a}(z)}{\sum_a e_a^2\,x f_1^a(x)\,D_1^{a}(z)} \;.
\ee

However, $H_1^{a}(z)$ is accessible in $e^+e^-\to \bar q q\to
2{\rm jets}$ where the quark transverse spin correlation induces 
a specific azimuthal correlation of two hadrons in opposite jets 
\cite{Boer:1997mf}
\be\label{Eq:A1-in-e+e}
    \di\sigma=\di\sigma_{\rm unp}\underbrace{\Biggl[
    1+\cos(2\phi_1)\frac{\sin^2\theta}{1+\cos^2\theta}
    \;C_{\rm G}\times
    \frac{\sum_a e_a^2 H_1^{a}H_1^{\bar a}}
    {\sum_a e_a^2 D_1^a D_1^{\bar a}}\Biggr]}_{\equiv A_1}
\ee
where $\phi_1$ is azimuthal angle of hadron 1 around z-axis along
hadron 2, and $\theta$ is electron polar angle. Also here we
assume the Gauss model and $C_{\rm
G}(z_1,z_2)=\frac{16}{\pi}{z_1z_2}/{(z_1^2+z_2^2)}$.

First experimental indications for the Collins effect were
obtained from studies of preliminary SMC data on SIDIS
\cite{Bravar:1999rq} and DELPHI data on charged hadron production
in $e^+e^-$ annihilations at the $Z^0$-pole \cite{Efremov:1998vd}.
More recently HERMES reported data on the Collins (SSA) in SIDIS
from proton target \cite{Airapetian:2004tw,Diefenthaler:2005gx}
giving the first unambiguous evidence that $H_1^a$ and $h_1^a(x)$
are non-zero, while in the COMPASS experiment
\cite{Alexakhin:2005iw} the Collins effect from a deuteron target
was found compatible with zero within error bars. Finally, last
year the BELLE collaboration presented data on sizeable azimuthal
correlation in $e^+e^-$ annihilations at a center of mass energy
of $60\,{\rm MeV}$ below the $\Upsilon$-resonance
\cite{Abe:2005zx,Ogawa:2006bm}.

The question which arises is: {\sl Are all these data from 
different SIDIS and $e^+e^-$ experiments compatible, i.e. due to 
the same effect, namely the Collins effect?}

In order to answer this question we extract $H_1^a$ from HERMES
\cite{Diefenthaler:2005gx} and BELLE  
\cite{Abe:2005zx,Ogawa:2006bm} data, and compare the obtained
ratios $H_1^a/D_1^a$ to each other and to other experiments. Such
``analyzing powers'' might be expected to be weakly
scale-dependent, as the experience with other spin observables
\cite{Ratcliffe:1982yj,Kotikov:1997df} indicates.

\subsection{Collins effect in SIDIS}

In order to extract information on Collins FF from SIDIS a model 
for the unknown $h_1^a(x)$ is needed. We use predictions from 
chiral quark-soliton model \cite{Schweitzer:2001sr} which 
provides a good description of unpolarized and helicity 
distribution \cite{Diakonov:1996sr}. On the basis of 
Eq.~(\ref{Eq:AUT-Collins-1}), the assumptions in 
Footnote~\ref{Footnote-1}, and the parameterizations 
\cite{Gluck:1998xa,Kretzer:2001pz} for $f_1^a(x)$ and $D_1^a(z)$ 
at $\la Q^2\ra=2.5\,{\rm GeV}^2$, we obtain from the HERMES data 
\cite{Diefenthaler:2005gx}:
\be\label{Eq:B-Gauss-H1perp12-fav}
    \la 2B_{\rm G}H_1^{\rm fav}\ra =  (3.5\pm 0.8)\;,\quad
    \la 2B_{\rm G}H_1^{\rm unf}\ra = -(3.8\pm 0.7)\;.
\ee
Here ``${\rm fav}$'' (``${\rm unf}$'') means favored $u\to\pi^+$ 
etc.\ (unfavored $u\to\pi^-$, etc.) fragmentation, and 
$\la\dots\ra$ denotes average over $z$ within the HERMES cuts 
$0.2\le z \le 0.7$.

\begin{figure}
\vspace{-5mm}
\includegraphics[width=1.3in]{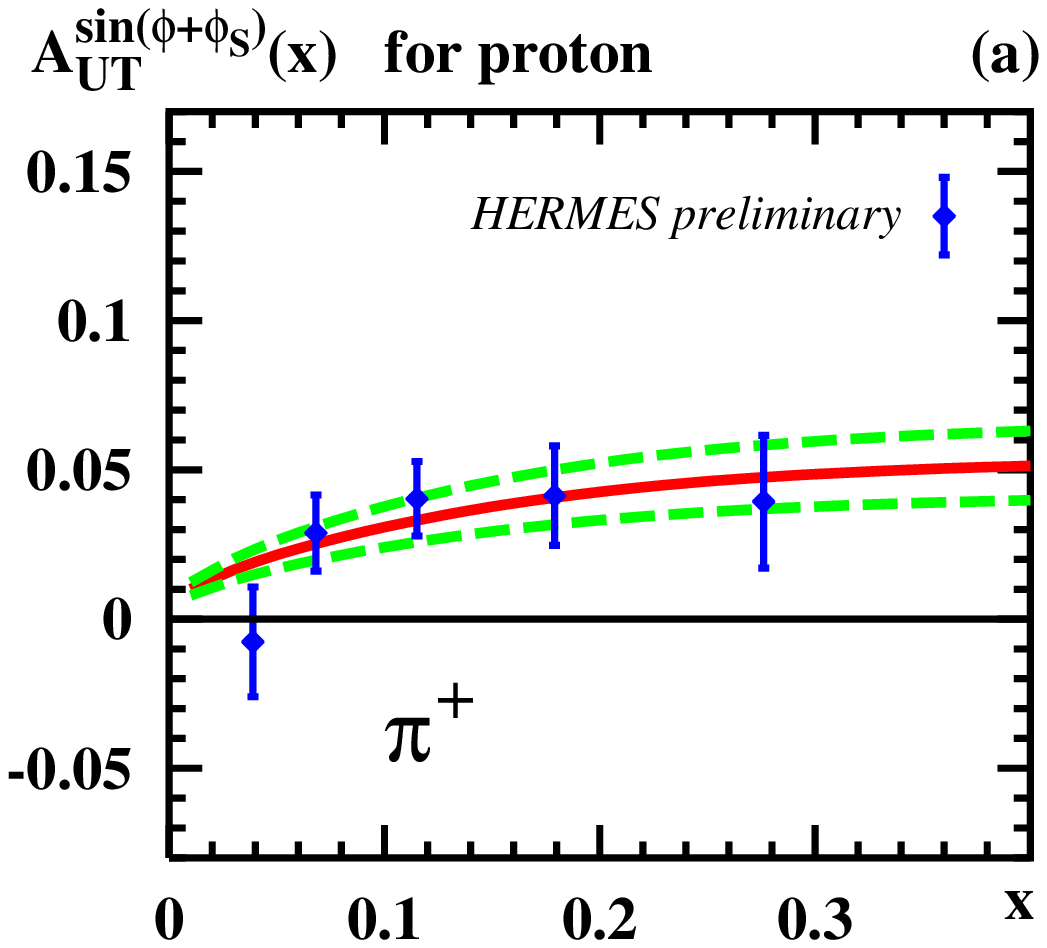}
\hspace{-3mm}
\includegraphics[width=1.3in]{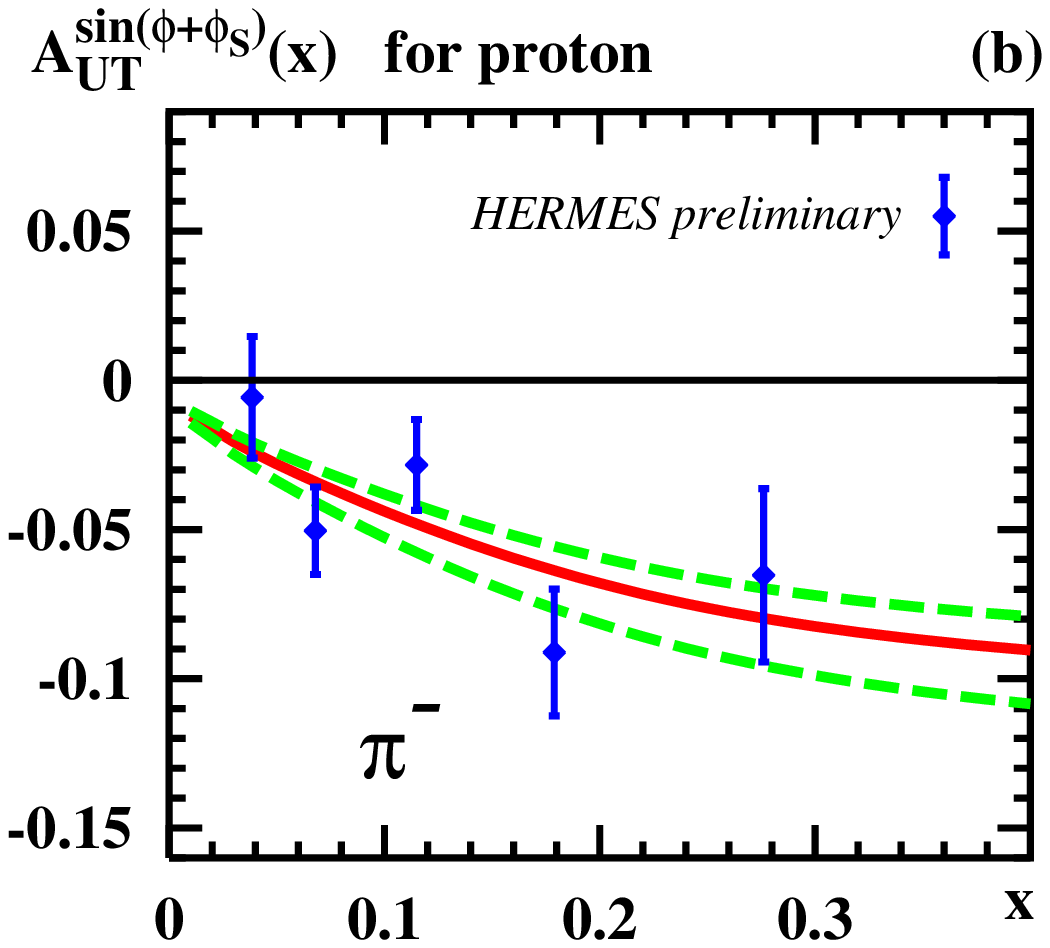}
\hspace{-6mm}
\includegraphics[width=1.4in]
{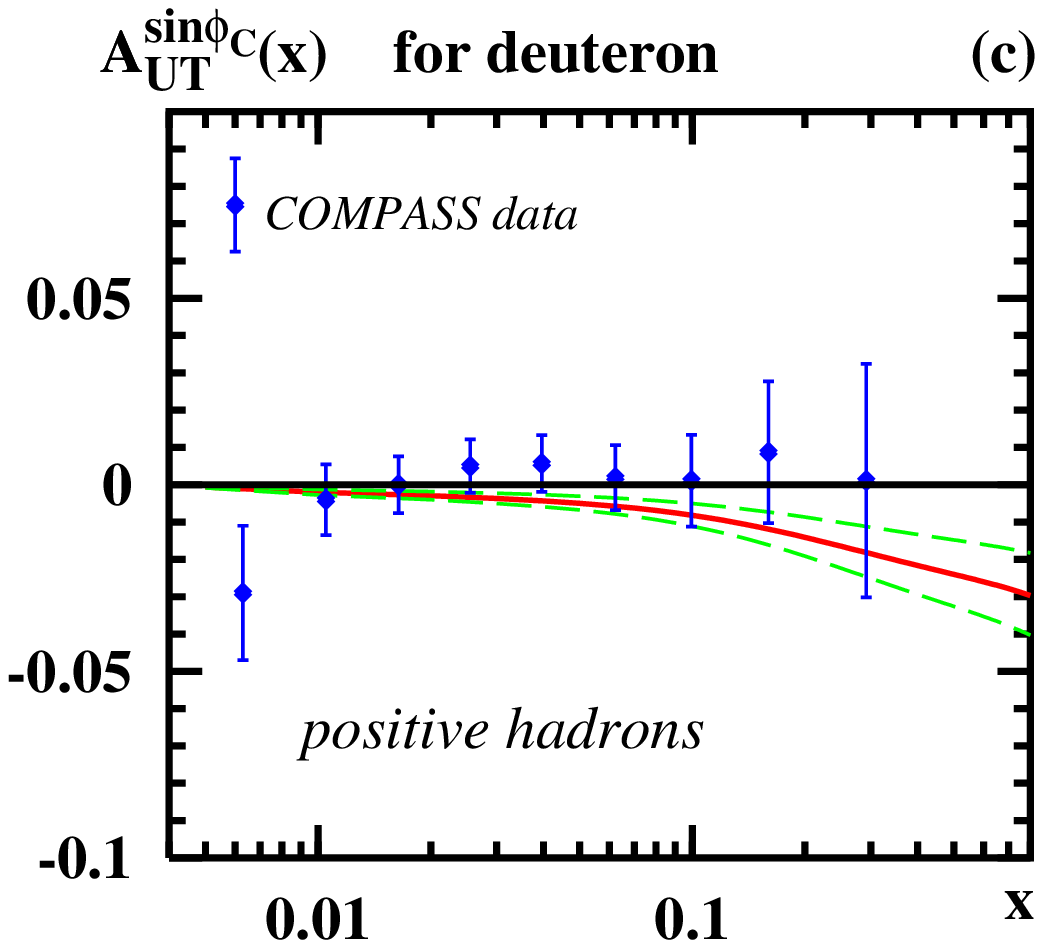} \hspace{-7mm}
\includegraphics[width=1.4in]
{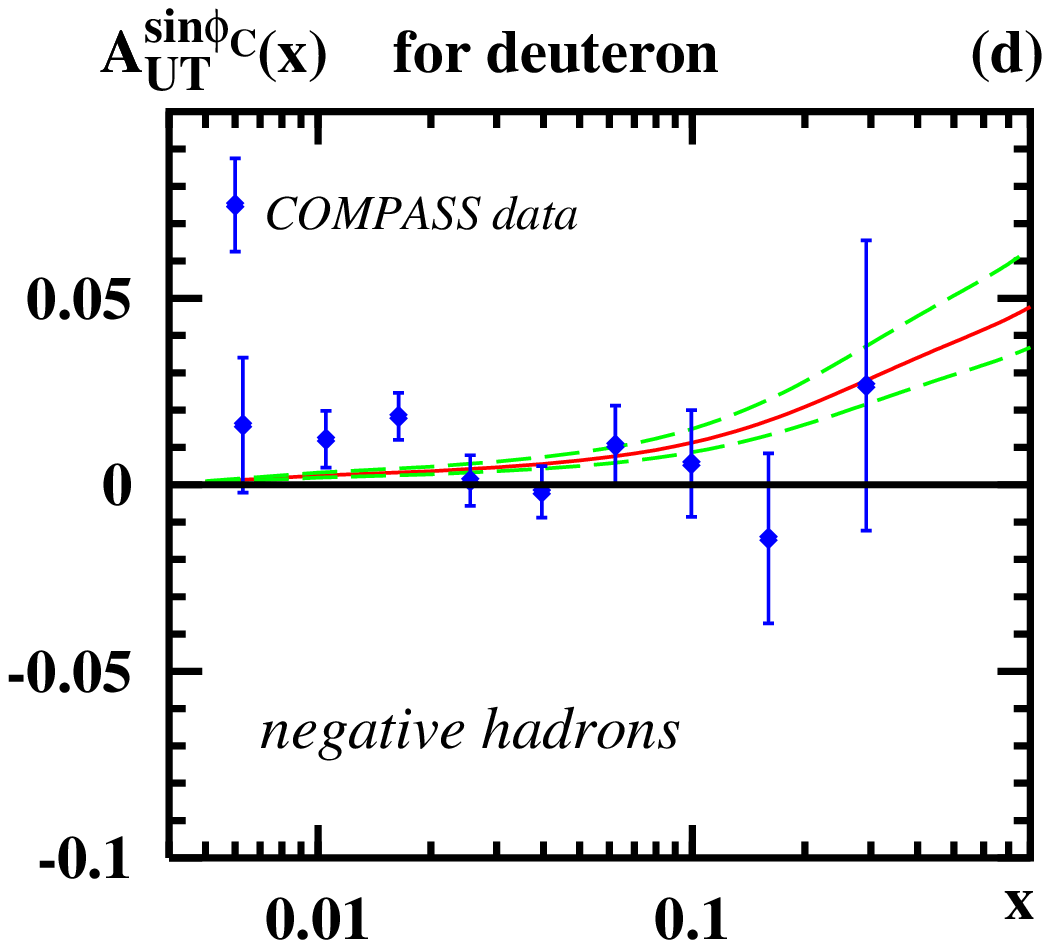}

\vspace{-3mm}
\caption{\label{Fig3:AUT-x}\footnotesize
Collins SSA $A_{UT}^{\sin(\phi+\phi_S)}$ as function of $x$ vs.\
HERMES \cite{Diefenthaler:2005gx} and new COMPASS
\cite{Alexakhin:2005iw} data.}
\end{figure}

Thus, the favored and unfavored Collins FFs appear to be of 
similar magnitude and opposite sign. The string fragmentation 
picture \cite{Artru:1995bh} and Sch\"afer-Teryaev sum rule 
\cite{Schafer:1999kn} provide a qualitative understanding of this 
behavior. The important role of unfavored FF becomes
more evident by considering the analyzing powers 
\be
\label{Eq:Apower-HERMES}
\frac{\la 2 B_{\rm G}H_1^{\rm fav}\ra}{\la D_1^{\rm fav}\ra}
\biggr|_{\rm HERMES}\hspace{-11mm}=(7.2\pm 1.7)\% \;, \;\;\;\; \\
\frac{\la 2 B_{\rm G}H_1^{\rm unf}\ra}{\la D_1^{\rm unf}\ra}
\biggr|_{\rm HERMES}\hspace{-11mm} = -(14.2\pm 2.7)\%\;.
\ee
Fit (\ref{Eq:B-Gauss-H1perp12-fav}) describes the HERMES proton
target data \cite{Diefenthaler:2005gx} on the Collins SSA (see
Figs.~\ref{Fig3:AUT-x}a, b) and is in agreement with COMPASS
deuteron data \cite{Alexakhin:2005iw} (Figs.~\ref{Fig3:AUT-x}c, 
d).

%

%
\begin{wrapfigure}[12]{HR}{0.38\textwidth}
\vspace{-12mm}
\begin{flushright}
\includegraphics[width=0.35\textwidth]{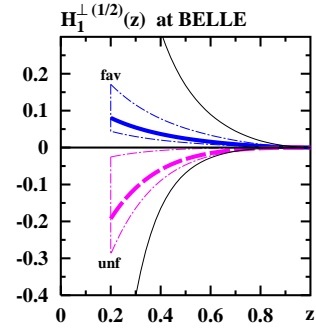}

\vspace{-5mm}\hfil
\begin{minipage}{0.35\textwidth}
\caption{\label{Fig5:BELLE-best-fit}\footnotesize
Collins FF $H_1^\aH(z)$ needed to explain
the BELLE data \cite{Abe:2005zx}.}
\end{minipage}
\end{flushright}
\end{wrapfigure}

\subsection{Collins effect in $\bf e^+e^-$}

The specific $\cos2\phi$ dependence of the cross section
(\ref{Eq:A1-in-e+e}) could arise also from hard gluon radiation or
detector acceptance effects. These effects, being flavor
independent, cancel out from the double ratio of $A_1^U$, where
both hadrons $h_1h_2$ are pions of unlike sign, to $A_1^L$, where
$h_1h_2$ are pions of like sign, i.e.
\be
\label{Eq:double-ratio}
\frac{A_1^U}{A_1^L}\approx 1 + \cos(2\phi_1)P_1(z_1,z_2)\;.
\ee

In order to describe the BELLE data \cite{Abe:2005zx} we
have chosen the Ansatz and obtained the best fit
\be
\label{Eq:best-fit-BELLE}
H_1^\aH(z) = C_a \,z\,D_1^a(z),\;\;\;
C_{\rm fav}=0.15,\;\;\;C_{\rm unf}=-0.45,
\ee
shown in Fig. \ref{Fig5:BELLE-best-fit} with 1-$\sigma$ error band
(the errors are correlated). Other Ans\"atze gave less
satisfactory fits.

%
\begin{figure}[b]
\begin{tabular}{cccc}
\vspace{-5mm} &&& \cr
\includegraphics[width=1.1in]{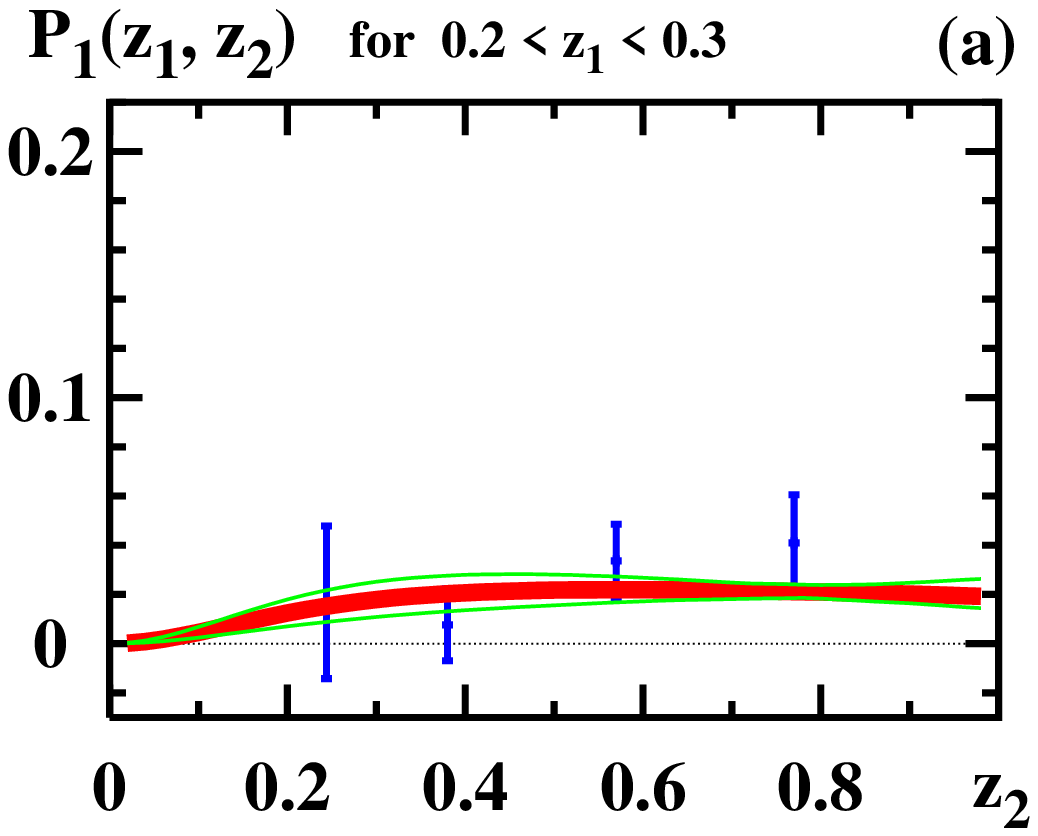}
&
\includegraphics[width=1.1in]{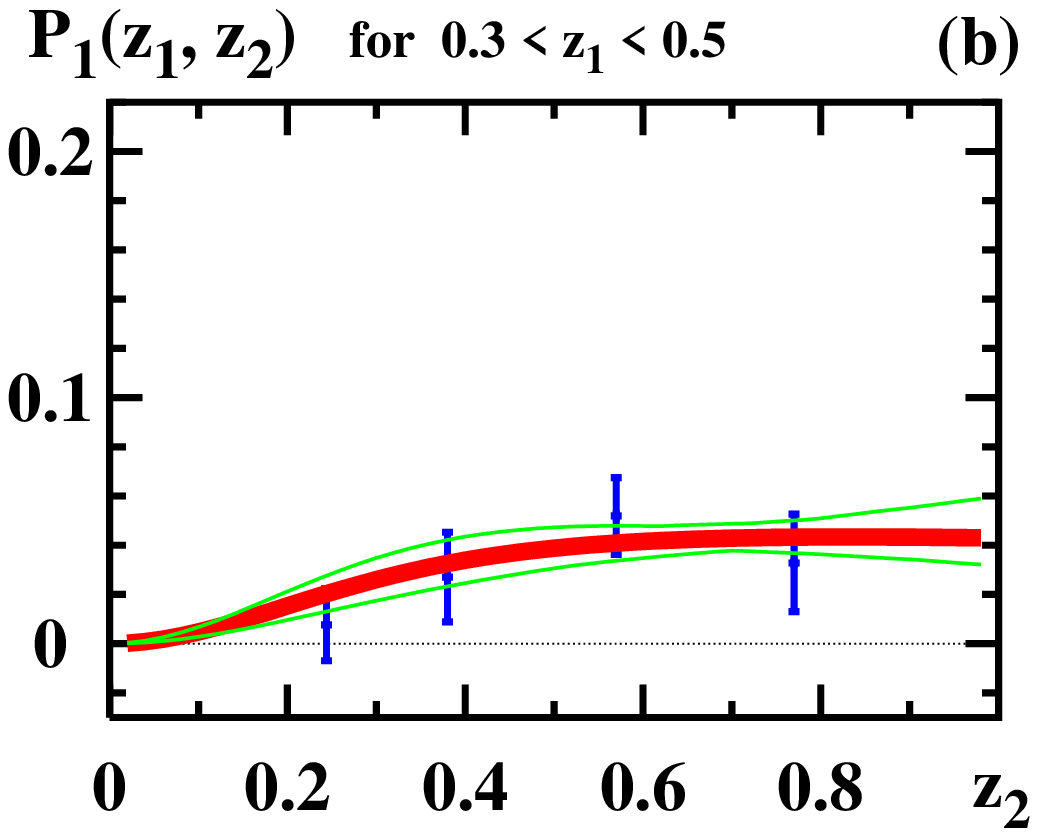}
&
\includegraphics[width=1.1in]{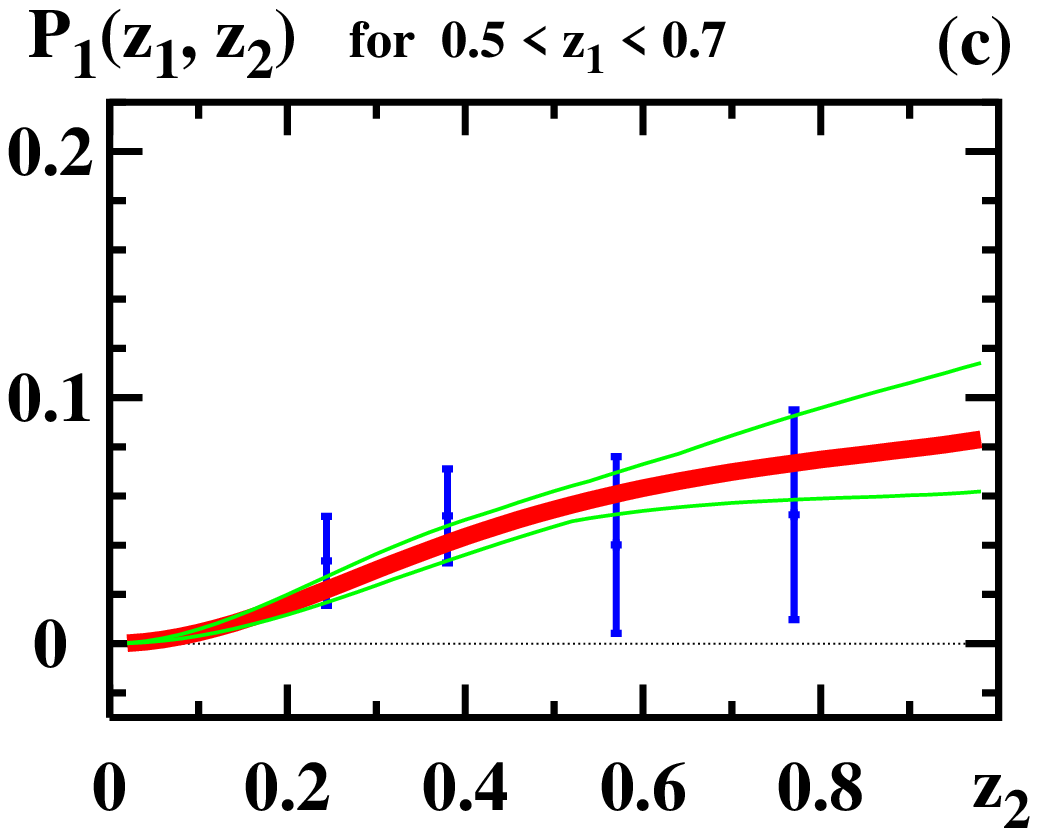}
&
\includegraphics[width=1.1in]{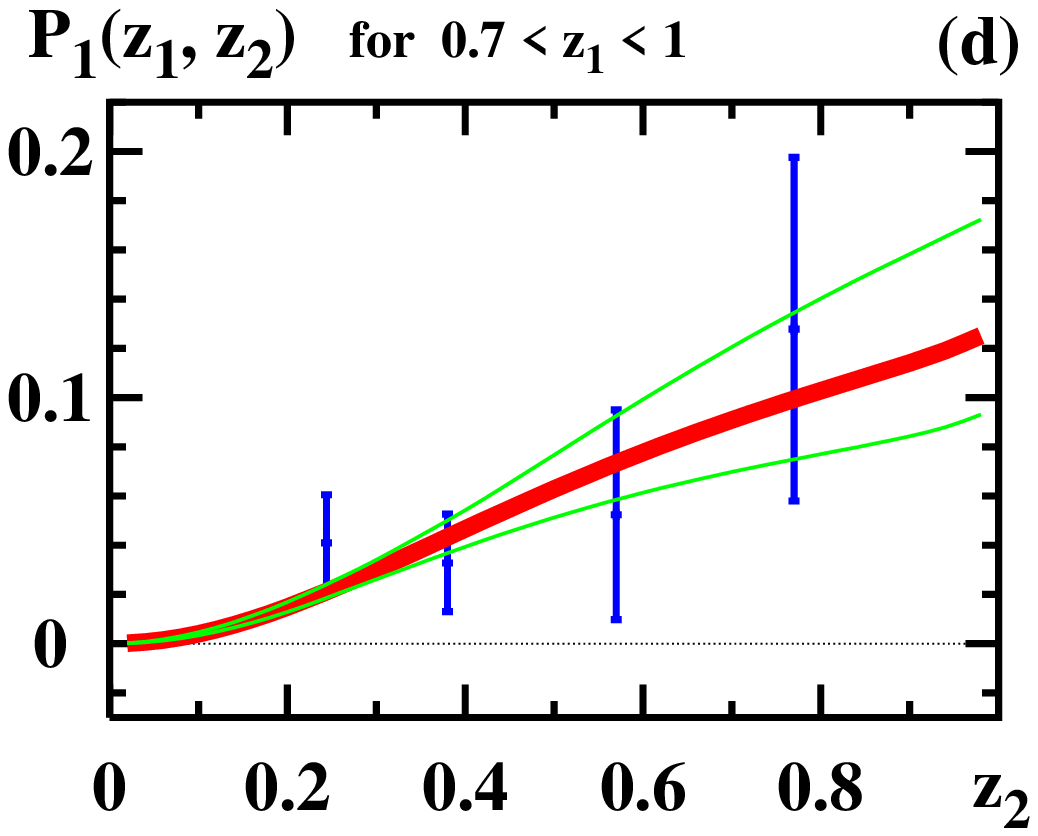}
\cr
\includegraphics[width=1.1in]{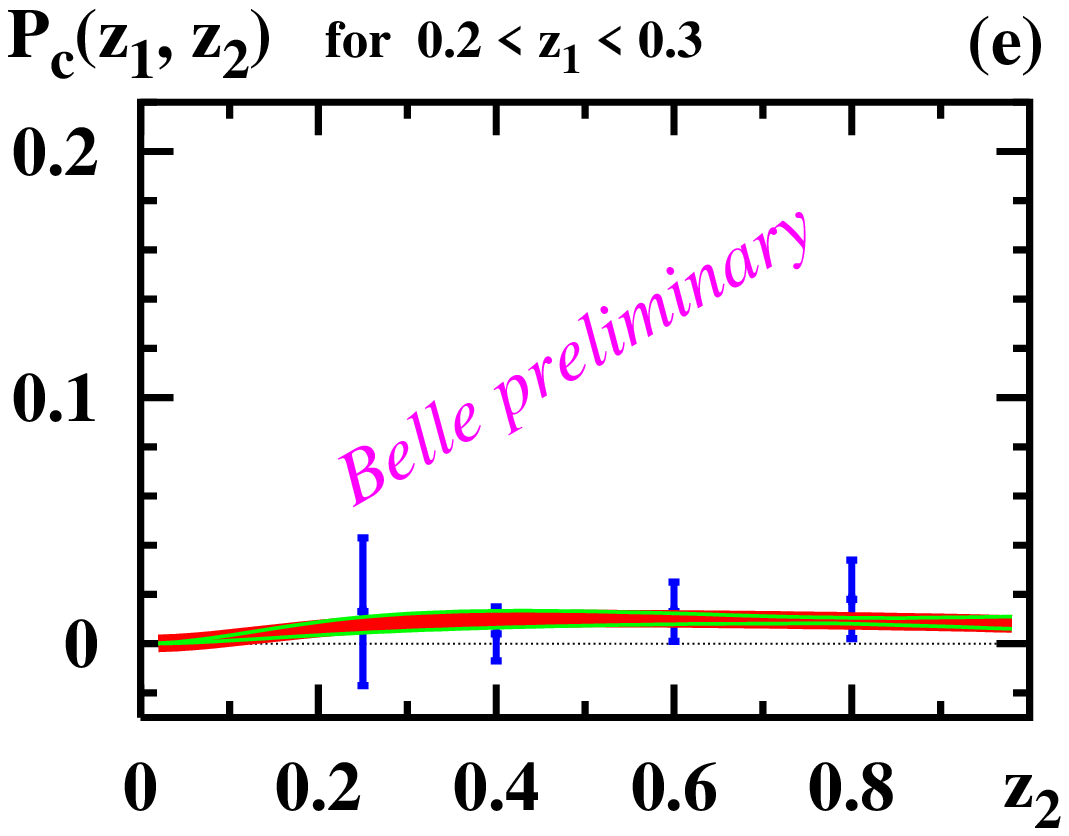}
&
\includegraphics[width=1.1in]{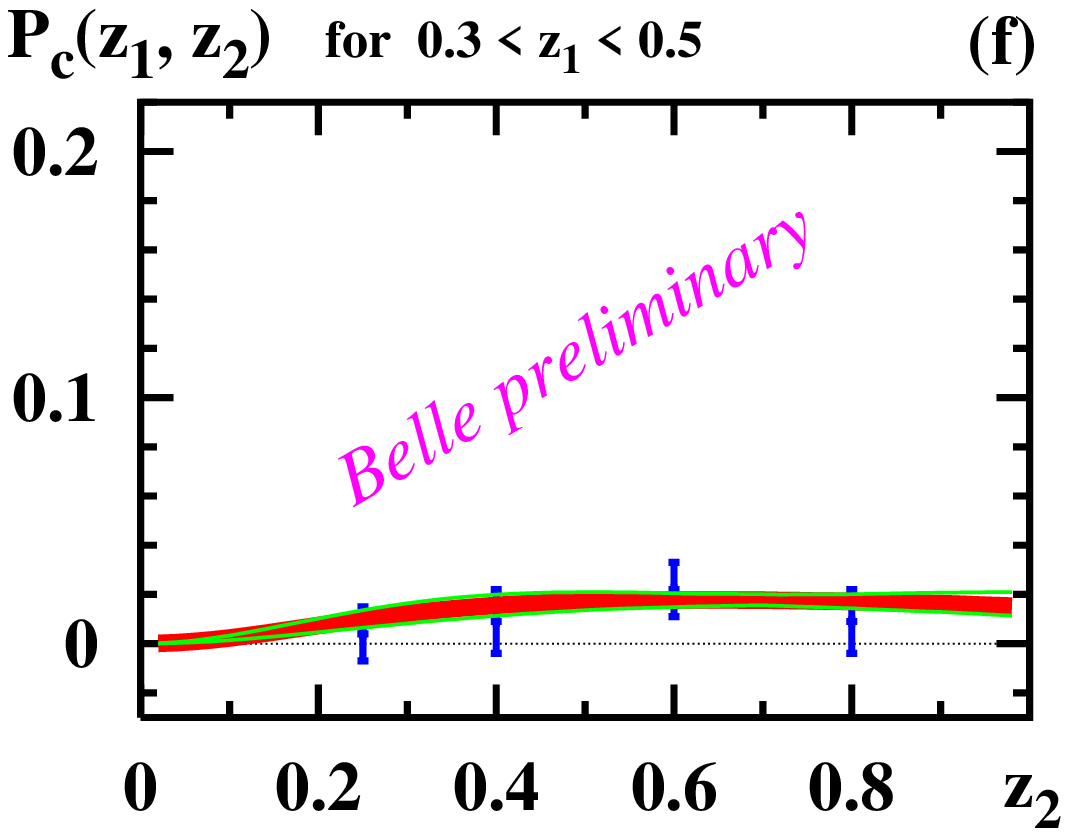}
&
\includegraphics[width=1.1in]{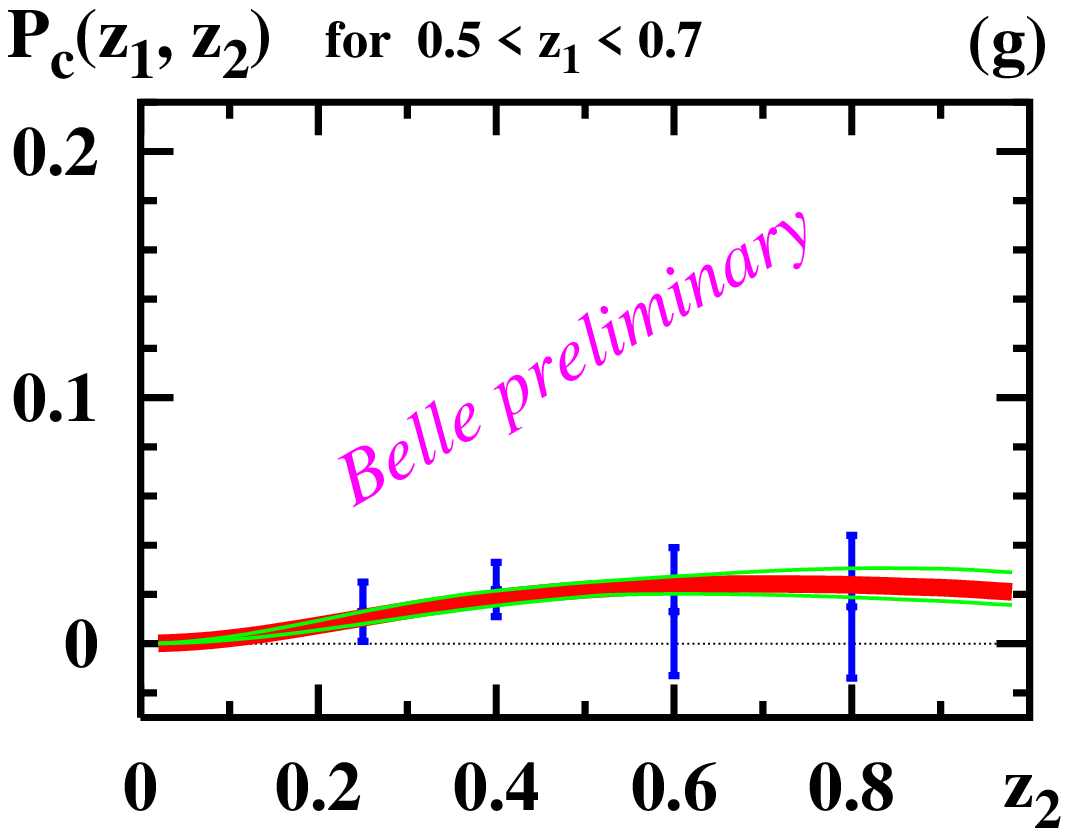}
&
\includegraphics[width=1.1in]{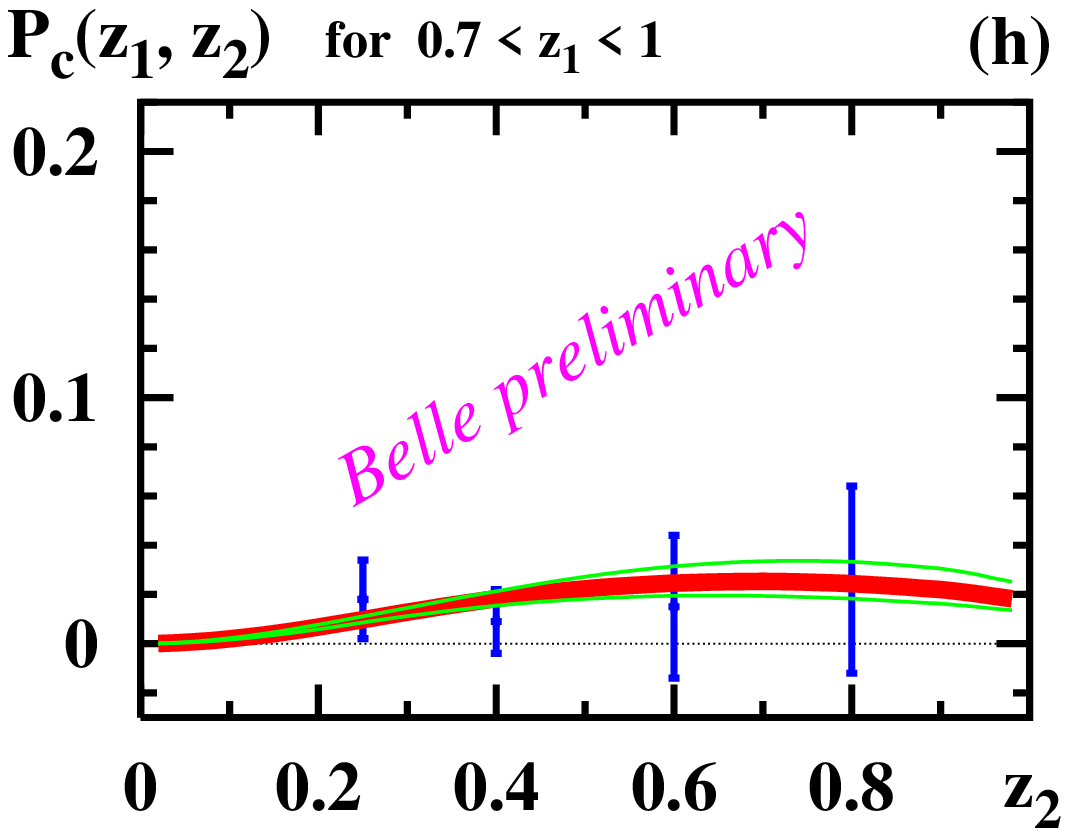}
\end{tabular}
\vspace{-5mm}
\caption{\label{Fig6:BELLE}\footnotesize
    {\bf a-d}:
    $\;\,P_1(z_1,z_2)$ as defined in
    Eq.~(\ref{Eq:double-ratio}) for fixed $z_1$-bins
    as function of $z_2$ vs.\ BELLE data \cite{Abe:2005zx}.
    {\bf e-h}:
    The observable $P_C(z_1,z_2)$ defined analogously, see text,
    vs. preliminary BELLE data reported in \cite{Ogawa:2006bm}.}
\end{figure}
%
Notice that azimuthal observables in $e^+e^-$-anni\-hilation are
bilinear in $H_1^\aH$ and therefore symmetric with respect to the
exchange of the signs of $H_1^{\rm fav}$ and $H_1^{\rm unf}$.
Thus in our Ansatz $P_1(z_1,z_2)$ is symmetric with respect to
the exchange ${\rm sign}(C_{\rm fav})\leftrightarrow{\rm
sign}(C_{\rm unf})$. (And not with respect to $C_{\rm
fav}\leftrightarrow C_{\rm unf}$ as incorrectly remarked in
\cite{Efremov:2006qm}.)

The BELLE data \cite{Abe:2005zx} unambiguously indicate that
$H_1^{\rm fav}$ and $H_1^{\rm unf}$ have opposite signs, but they
cannot tell us which is positive and which is negative. The
definite signs in (\ref{Eq:best-fit-BELLE}) and Fig.
\ref{Fig5:BELLE-best-fit} are dictated by SIDIS data
\cite{Diefenthaler:2005gx} (and our model \cite{Schweitzer:2001sr}
with $h_1^u(x)>0$, see Sect. 3.1).

In Fig.~\ref{Fig6:BELLE}a-d the BELLE data \cite{Abe:2005zx} are
compared to the theoretical result for $P_1(z_1,z_2)$ obtained on
the basis of the best fit shown in Fig.~\ref{Fig5:BELLE-best-fit}.

Most interesting recent news are the preliminary BELLE data
\cite{Ogawa:2006bm} for the ratio of azimuthal asymmetries of
unlike sign pion pairs, $A_1^U$, to all charged pion pairs,
$A_1^C$. The new observable $P_C$ is defined analogously to $P_1$
in Eq.~(\ref{Eq:double-ratio}) as $A_1^U/A_1^C \approx
(1+\cos(2\phi)\,P_C)$. The fit (\ref{Eq:best-fit-BELLE}) ideally
describes the new experimental points (see
Figs.~\ref{Fig6:BELLE}e-h)!

\subsection{BELLE vs.~HERMES}

In order to compare Collins effect in SIDIS at HERMES
\cite{Airapetian:2004tw,Diefenthaler:2005gx} and in
$e^+e^-$-annihilation at BELLE \cite{Abe:2005zx} we consider the
ratios $H_1^a/D_1^a$ which might be less scale dependent. The
BELLE fit in Fig.~\ref{Fig5:BELLE-best-fit} yields in the HERMES
$z$-range:
\be\label{Eq:Apower-BELLE}
    \frac{\la 2H_1^{\rm fav}\ra}{\la D_1^{\rm fav}\ra}
    \biggr|_{\rm BELLE}\hspace{-9mm} = (5.3\cdots 20.4)\%,\quad
    \frac{\la 2H_1^{\rm unf}\ra}{\la D_1^{\rm
    unf}\ra} \biggr|_{\rm BELLE}\hspace{-9mm} =
    -(3.7\;\cdots\;41.4)\%  \;.
\ee

%
\begin{wrapfigure}[12]{R}{2.8in}
\vspace{-10mm}
\begin{flushright}
\includegraphics[width=1.25in]{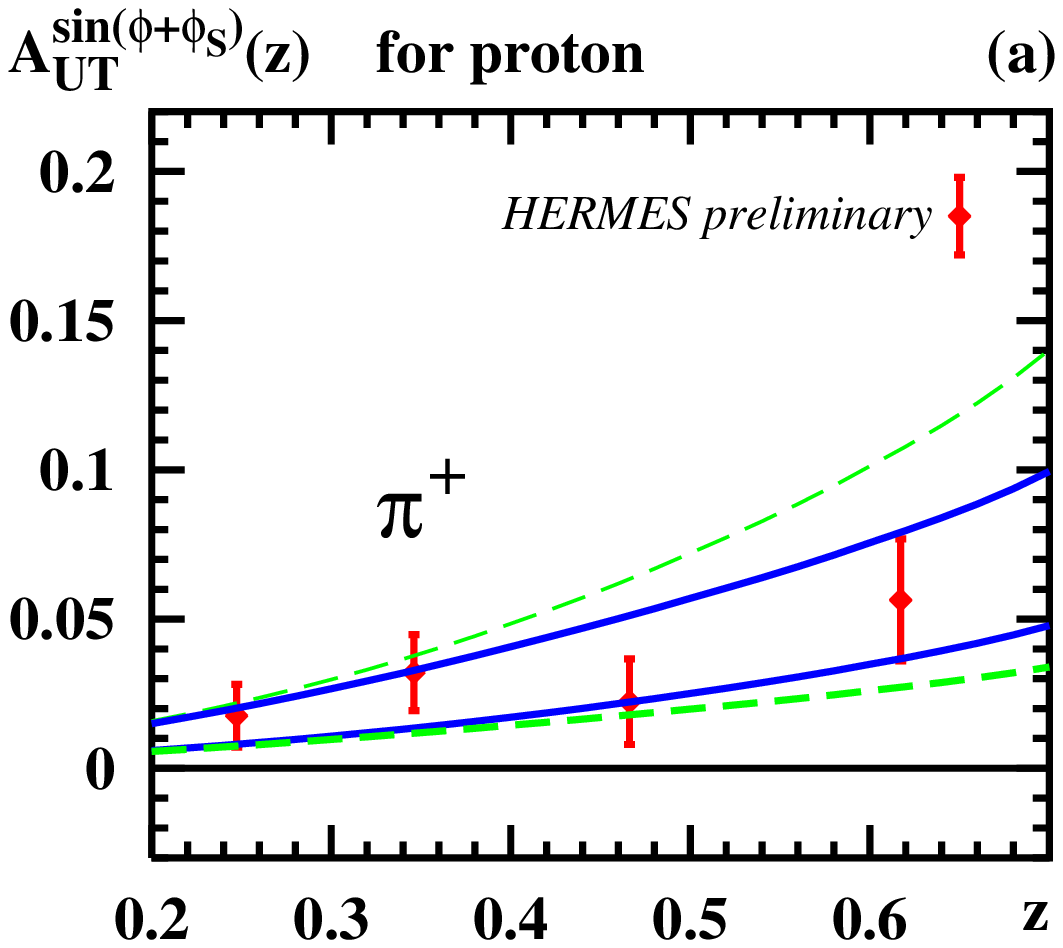}
\includegraphics[width=1.25in]{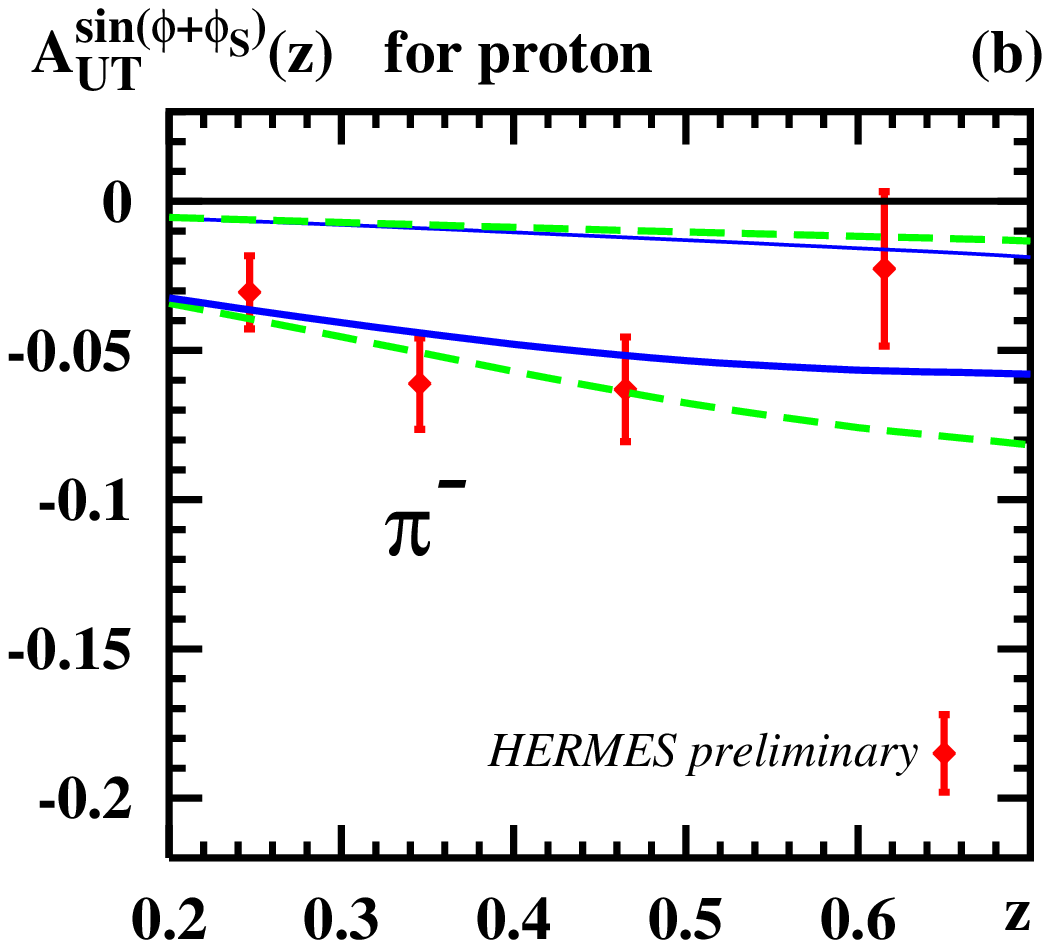}
\end{flushright}

\vspace{-12mm}
\begin{flushright}
\begin{minipage}{2.5in}
\caption{\label{Fig7:HERMES-AUT-z-from-BELLE} \footnotesize
    The Collins SSA $A_{UT}^{\sin(\phi+\phi_S)}(z)$ as function
    of $z$. The theoretical curves are based on the fit of
    $H_1^a(z)$ to the BELLE data under the assumption
    (\ref{Eq:assume-weak-scale-dep}).
    The dashed lines indicate
    the sensitivity of the SSA to the unknown ratio of the
    Gaussian widths.
    }
\end{minipage}
\end{flushright}
\end{wrapfigure}
%

Comparing the above numbers (the errors are correlated!) to the
result in Eq.~(\ref{Eq:Apower-HERMES}) we see that the effects at
HERMES and at BELLE are compatible. The central values of the
BELLE analyzing powers seem to be systematically larger but this
could partly be attributed to evolution effects and to the factor
$B_{\rm G}<1$ in Eq.~(\ref{Eq:Apower-HERMES}).

By assuming a weak scale-dependence also for the $z$-dependent ratios
\be\label{Eq:assume-weak-scale-dep}
    \frac{H_1^\aH(z)}{D_1^a(z)}\biggr|_{\rm BELLE}
    \approx\;
    \frac{H_1^\aH(z)}{D_1^a(z)}\biggr|_{\rm HERMES}
\ee
and considering the 1-$\sigma$ uncertainty of the BELLE fit in
Fig. \ref{Fig5:BELLE-best-fit} and the sensitivity to unknown
Gaussian widths of $H_1^a(z)$ and $h_1^a(x)$, c.f.\  Footnote~1
and Ref.~\cite{Efremov:2006qm}, one obtains also a satisfactory
description of the $z$-dependence of the SIDIS HERMES data
\cite{Diefenthaler:2005gx}, see
Fig.~\ref{Fig7:HERMES-AUT-z-from-BELLE}.

These observations allow --- within the accuracy of the first data
and the uncertainties of our study --- to draw the conclusion that
it is, in fact, the same Collins effect at work in SIDIS
\cite{Airapetian:2004tw,Diefenthaler:2005gx,Alexakhin:2005iw}
and in $e^+e^-$-annihilation \cite{Abe:2005zx,Ogawa:2006bm}.
Estimates indicate that the early preliminary DELPHI result
\cite{Efremov:1998vd} is compatible with
these findings,
see \cite{Efremov:2006qm} for details.

%
\begin{wrapfigure}[12]{HR}{2.1in}
\vspace{-5mm}
\centerline{\includegraphics[width=1.7in]{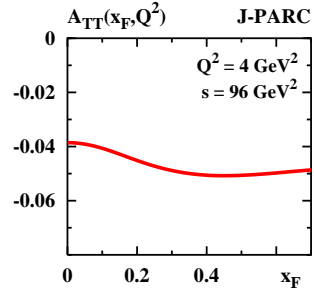}}

\vspace{-5mm}\hfil
\begin{minipage}{1.9in}
\caption{\label{Fig-DY-ATT-pp-JPARC}\footnotesize
Double spin asymmetry $A_{TT}$ in DY, Eq.~(\ref{Eq:DY}), vs.\
$x_F$ for the kinematics of J-PARC from \cite{in-progress}.}
\end{minipage}
\end{wrapfigure}
\subsection{Transversity in Drell-Yan process}
The double-spin asymmetry observable in Drell-Yan (DY) lepton-pair
production in proton-proton ($pp$) collisions is given in LO by
\be\label{Eq:DY}
    A_{TT}(x_F) =
    \frac{\sum_a e_a^2 h_1^a(x_1) h_1^{\bar a}(x_2)}
         {\sum_a e_a^2 f_1^a(x_1) f_1^{\bar a}(x_2)}
\ee
where $x_F=x_1-x_2$ and $x_1x_2=\frac{Q^2}{s}$. In the kinematics
of RHIC $A_{TT}$ is small and difficult to measure
\cite{Bunce:2000uv}.

In the J-PARC experiment with $E_{\rm beam}=50\,{\rm GeV}$
$A_{TT}$ would reach $-5\,\%$ in the model
\cite{Schweitzer:2001sr}, see Fig. \ref{Fig-DY-ATT-pp-JPARC}, and
could be measured \cite{J-PARC-proposal}. The situation is
similarly promising in proposed U70-experiment
\cite{Abramov:2005mk}.

Finally, in the PAX-experiment proposed at GSI \cite{PAX}
in polarized $\bar pp$ collisions one may expect
$A_{TT}\sim(30\cdots50)\%$ \cite{Efremov:2004qs}.
There $A_{TT}\propto h_1^u(x_1) h_1^u(x_2)$ to a good approximation,
due to $u$-quark ($\bar u$-quark) dominance in the proton (anti-proton)
\cite{Efremov:2004qs}.

\section{Conclusions}
Within the uncertainties of our study we find that the SIDIS data
from HERMES \cite{Airapetian:2004tw,Diefenthaler:2005gx} and
COMPASS \cite{Alexakhin:2005iw} on the Sivers and Collins SSA from
different targets are in agreement with each other and with BELLE
data on azimuthal correlations in $e^+e^-$-annihilations
\cite{Abe:2005zx}.

At the present stage large-$N_c$ predictions give useful 
constraints and are compatible with data. 

Data on kaon production provide new interesting information on 
sea Sivers-$\bar q$.

For the Collins FF the following picture emerges: favored and
unfavored Collins FFs appear to be of comparable magnitude but
have opposite signs, and $h_1^u(x)$ seems close to saturating the
Soffer bound while the other $h_1^a(x)$ are presently
unconstrained \cite{Efremov:2006qm}.

These findings are in agreement with old DELPHI
\cite{Efremov:1998vd} and with the most recent BELLE data
\cite{Ogawa:2006bm} and with independent theoretical studies
\cite{Vogelsang:2005cs}. Further data from SIDIS (COMPASS, JLAB 
\cite{Chen:2005dq}, HERMES) and $e^+e^-$ colliders (BELLE) will 
help to refine and improve this first picture.

The understanding of the novel functions $f_1^{\perp a},\,h_1^a$
and $H_1^a(z)$ emerging from SIDIS and $e^+e^-$-annihilations,
however, will be completed  only thanks to future data on double 
transverse spin asymmetries in the Drell-Yan process. Experiments 
are in progress or planned at RHIC, J-PARC, COMPASS, U70 and PAX 
at GSI.


\section*{Acknowledgments}
This work is  supported by BMBF (Verbundforschung), COSY-J\"ulich
project, the Transregio Bonn-Bochum-Giessen, and is part of the by
EIIIHT project under contract number RII3-CT-2004-506078. A.E. is
also supported by RFBR grant 06-02-16215, by RF MSE
RNP.2.2.2.2.6546 (MIREA) and by the Heisenberg-Landau Program of
JINR.


\end{document}